\documentstyle[prl,eqsecnum,aps]{revtex}
\twocolumn

\begin{document}
\author{Jian Qi Shen \footnote{E-mail address: jqshen@coer.zju.edu.cn, jqshencn@yahoo.com.cn}}
\address{Zhejiang Institute of Modern Physics and Department of
Physics, Zhejiang University (Yuquan Campus), \\ Hangzhou 310027,
People's Republic of China}
\date{\today }
\title{The supersymmetric Pegg-Barnett oscillator}
\maketitle

\begin{abstract}
The su$(n)$ Lie algebraic structure of the Pegg-Barnett oscillator
that possesses a finite-dimensional number-state space is
demonstrated. The supersymmetric generalization of Pegg-Barnett
oscillator is suggested. It is shown that such a supersymmetric
Pegg-Barnett oscillator may have some potential applications, {\it
e.g.}, the mass spectrum of the charged leptons.

PACS Ref: 03.65.Fd;  02.20.Sv

Keywords:  Pegg-Barnett oscillator; su($n$) algebraic
 structures
\end{abstract}
\pacs{}
\section{Introduction}
The quantum harmonic oscillator possessing an infinite-dimensional
number-state space ({\it i.e.}, the maximum occupation number $s$
tends to infinity) can well model the Bosonic fields. For example,
light is considered a set of infinite number of harmonic
oscillators. Since the classical observable phase of light has no
corresponding Hermitian operator counterpart (quantum optical
phase)\cite{Louisell,Susskind,Carruthers}, we will meet with some
difficulties when we investigate the number-phase uncertainty
relations of the maser and squeezed state in quantum optics. These
difficulties may be as follows: (i) the exponential-form operator
$\exp [i\hat{\phi}]$ is {\it not} unitary, where $\hat{\phi}$ is a
phase operator; (ii) the number-state expectation value of Dirac's
quantum relation $[\hat{\phi}, \hat{N}]=-i$ is zero, {\it i.e.},
$\langle n |[\hat{\phi}, \hat{N}]| n\rangle=0$. Here, $\hat{N}$
denotes the occupation-number operator of photon fields. The
expectation value of the right-handed side of Dirac's quantum
relation, {\it i.e.}, $\langle n |(-i)| n\rangle$ is, however,
nonvanishing; (iii) the number-phase uncertainty relation $\Delta
N\Delta \phi\geq \frac{1}{2}$ would imply that a well-defined
number state would actually have a phase uncertainty of greater
than $2 \pi$\cite{Pegg}. In order to obtain a well-behaved
Hermitian phase operator, Pegg and Barnett defined a phase state
as
$|\theta\rangle=\lim_{s\rightarrow\infty}(s+1)^{-\frac{1}{2}}\sum^{s}_{n=0}\exp
(in\theta)|n\rangle$, where $|n\rangle$ ($n=0, 1, 2,..., s$) are
the $s+1$ number states, which span an $(s+1)$-dimensional state
space, and thus suggested an alternative, and physically
indistinguishable, mathematical model of the monomode field
corresponding to a finite but arbitrarily large state
space\cite{Pegg}. This means that the state space $\{|n\rangle\}$
with $0 \leq n\leq s$ has a finitely upper level ($|s\rangle$) and
the maximum occupation number of particles is $s$ rather than
infinity. It was shown that Pegg-Barnett (P-B) approach has
several advantages over the conventional Susskind-Glogower
formulation\cite{Susskind}. For example, the Pegg-Barnett phase
operator is consistent with the vacuum being a state of random
phase, while the Susskind-Glogower phase operator does not
demonstrate such a property of the vacuum\cite{Pegg}. The
resulting number-phase commutator in the P-B approach does not yet
lead to any inconsistencies, but satisfies the condition for
Poisson-bracket-commutator correspondence. The P-B formulation
opened up new opportunities for treating some problems in quantum
optics ({\it e.g.}, atomic coherent population trapping (CPT) and
electromagnetically induced transparency (EIT)\cite{Purdy}) and in
quantum information\cite{Koniorczyk}.

In this paper we will further consider the Pegg-Barnett harmonic
oscillator that involves a finitely large state space, and show
that it possesses an su($n$) Lie algebraic structures. Based on
this consideration, we will generalize the Pegg-Barnett oscillator
to a supersymmetric case.

\section{Algebraic structure of P-B oscillator}
In this section, we will study the algebraic structures of P-B
harmonic oscillators with arbitrary occupation number $s$. For the
Pegg-Barnett (P-B) harmonic oscillator with a finite but
arbitrarily large state space of $s+1$ dimensions, the commutation
relation between the annihilation and creation operators is $[a,
a^{\dagger}]={\mathcal A}$ instead of the conventional commutation
relation ($[a, a^{\dagger}]={\mathcal I}$) of the Bosonic field.
This means that the non-semisimple Lie algebra should be
generalized to a semisimple one. It can be verified that the
matrix representation of the operators $a$, $a^{\dagger}$ and
${\mathcal A}$ takes the following form (in the number-state
base-vector set)
\begin{eqnarray}
& &  a_{mn}=\sqrt{n}\delta_{m, n-1},  \quad
a^{\dagger}_{mn}=\sqrt{n+1}\delta_{m, n+1},    \nonumber \\
& &    {\mathcal A}_{mn}=\delta_{mn}-(s+1)\delta_{ms}\delta_{ns},
\end{eqnarray}
where the subscript $m, n$ (which run from $0$ to $s$ only) denote
the matrix row-column indices. The remaining generators ${\mathcal
M}, {\mathcal M}^{\dagger}, {\mathcal K}, {\mathcal F}, {\mathcal
F}^{\dagger}, ...$ can be obtained as follows ($0\leq m, n\leq
s$):
\begin{eqnarray}
& &       \left[a, {\mathcal
A}\right]_{mn}=(s+1)\sqrt{s}(-\delta_{m+1, s}\delta_{n
s})=(s+1)\sqrt{s}{\mathcal M}_{mn},   \nonumber \\
& &      \left[a^{\dagger}, {\mathcal
A}\right]_{mn}=-(s+1)\sqrt{s}(-\delta_{m s}\delta_{n+1,
s})=-(s+1)\sqrt{s}{\mathcal M}^{\dagger}_{mn},
              \nonumber \\
& &      \left[{\mathcal M}, {\mathcal
M}^{\dagger}\right]_{mn}=-(\delta_{ms}\delta_{ns}-\delta_{m+1,s}\delta_{n+1,s})=-{\mathcal
K}_{mn},
\nonumber \\
& &      \left[ {\mathcal A}, {\mathcal M}\right]=(1+s){\mathcal
M}, \quad \left[ {\mathcal A}, {\mathcal
M}^{\dagger}\right]=-(1+s){\mathcal M}^{\dagger},
\nonumber \\
& &      \left[a, {\mathcal M}\right]_{mn}=-\sqrt{s-1}\delta_{m+1,
s-1}\delta_{ns}=-\sqrt{s-1}{\mathcal F}_{mn},
\nonumber \\
& &      \left[a^{\dagger}, {\mathcal
M}^{\dagger}\right]_{mn}=\sqrt{s-1}\delta_{ms}\delta_{n+1,
s-1}=\sqrt{s-1}{\mathcal F}^{\dagger}_{mn}, \nonumber  \\
& &      \left[{\mathcal K}, {\mathcal F}\right]=-{\mathcal F},
\quad \left[{\mathcal K}, {\mathcal F}^{\dagger}\right]={\mathcal
F}^{\dagger},   \nonumber  \\
& &       \left[{\mathcal M}, {\mathcal K}\right]=2{\mathcal M},
\quad             \left[{\mathcal M}^{\dagger}, {\mathcal
K}\right]=-2{\mathcal M}^{\dagger}, \quad ...
\label{eq4}
\end{eqnarray}

In the following, we will take into account the particular cases
of $s=1, 2$. According to the commutation relations (\ref{eq4})
for the case of $s=1$, in the number-state space, the matrix
representation of the generators $a, a^{\dagger} $ and $ {\mathcal
A}$ are reduced to $a=({\sigma_{1}+i\sigma_{2}})/{2}$,
$a^{\dagger}=({\sigma_{1}-i\sigma_{2}})/{2}$ and ${\mathcal
A}=\sigma_{3}$, where $\sigma_{i}$'s ($i=1,2,3$) are Pauli's
matrices. In the meanwhile, the rest generators ${\mathcal M},
{\mathcal M}^{\dagger}, {\mathcal K}, {\mathcal F}, {\mathcal
F}^{\dagger}, ...$ are ${\mathcal M}=-a, {\mathcal
M}^{\dagger}=-a^{\dagger}, {\mathcal K}=-{\mathcal A}, {\mathcal
F}={\mathcal O}, {\mathcal F}^{\dagger}={\mathcal O}, ...$,
respectively. It is thus shown that the three generators $\{a,
a^{\dagger}, {\mathcal A}\}$ form a Lie algebra. It follows that
$aa^{\dagger}+a^{\dagger}a={\mathcal I}$. Since the algebraic
generators of su(2) Lie algebra can be constructed in terms of
Pauli's matrices, the P-B harmonic oscillator with $s=1$
corresponds to the Fermionic fields and possesses an su($2$) Lie
algebraic structure.

For the case of $s=2$, it can also be shown that the algebraic
generators $a, a^{\dagger}, {\mathcal A}, {\mathcal M}, {\mathcal
M}^{\dagger}, {\mathcal K}, {\mathcal F}, {\mathcal F}^{\dagger}$
form an sl($3$) algebra. The eight Gell-Mann matrices,
$\lambda_{i}$, can therefore be constructed in terms of these
algebraic generators, {\it i.e.},
\begin{eqnarray}
& & \lambda_{1}=a+a^{\dagger}+\sqrt{2}({\mathcal M}+{\mathcal
M}^{\dagger}),  \nonumber \\
& & \lambda_{2}=i[a^{\dagger}-a+\sqrt{2}({\mathcal
M}^{\dagger}-{\mathcal M})],   \nonumber \\
& &      \lambda_{3}={\mathcal A}+2{\mathcal K},  \quad
\lambda_{4}={\mathcal F}+{\mathcal F}^{\dagger},
                 \nonumber \\
& &    \lambda_{5}=i({\mathcal F}^{\dagger}-{\mathcal F}),   \quad
\lambda_{6}=-({\mathcal M}+{\mathcal M}^{\dagger}), \nonumber \\
& &    \lambda_{7}=-i({\mathcal M}^{\dagger}-{\mathcal M}), \quad
\lambda_{8}=\frac{1}{\sqrt{3}}\lambda_{8}.
\end{eqnarray}
Thus we demonstrated that the P-B harmonic oscillator with $s=2$
possesses an su($3$) Lie algebraic structure.

As to the case of arbitrary integer $s$, it can be verified that
the P-B harmonic oscillator possesses an su($s+1$) Lie algebraic
structure: if ${\mathcal G}$ represents the linear combination of
the Hermitian operators, and consequently ${\mathcal G}={\mathcal
G}^{\dagger}$, then the exponential-form group element operator
$U=\exp (i{\mathcal G})$ is unitary. Besides, since $a$,
$a^{\dagger}$ and ${\mathcal A}$ are traceless, all the generators
derived by the commutators in (\ref{eq4}) (and hence ${\mathcal
G}$) are also traceless due to the cyclic invariance in the trace
of matrices product. Thus the determinant of the group element $U$
is unity, {\it i.e.}, ${\rm det}U=1$, because of ${\rm det}U=\exp
[{\rm tr}(i{\mathcal G})]$. Since it is known that such a group
element $U$ that satisfies simultaneously the above two conditions
is the group element of the su($n$) Lie group, the
high-dimensional Gell-Mann matrices, which closes the
corresponding su($n$) Lie algebraic commutation relations among
themselves, can also be constructed in terms of the generators $a,
a^{\dagger}, {\mathcal A}, {\mathcal M}, {\mathcal M}^{\dagger},
{\mathcal K}, {\mathcal F}, {\mathcal F}^{\dagger}, ...$ presented
above. It is thus concluded that the P-B harmonic oscillator with
a maximum occupation number $s$ has an $(s+1)$-dimensional
number-state space and possesses an su($s+1$) Lie algebraic
structure.

Considering the case of $s\rightarrow \infty$ is of typically
physical interest. Apparently, it is seen that ${\mathcal A}$
tends to a unit matrix ${\mathcal I}$, and all the remaining
generators (except $a$ and $ a^{\dagger}$), the off-diagonal
matrix elements of which approach zero, are thus reduced to
${\mathcal O}$. This, therefore, means that the P-B harmonic
oscillator with an infinite-dimensional state space just
corresponds to the Bosonic fields.

We hope the consideration of algebraic structures of P-B
oscillator presented here might be applicable to the investigation
of some related topics such as fractional statistics,
anyon\cite{Wilczek} and cyclic representation of quantum algebra
(group)\cite{Fujikawa}.

\section{The supersymmetric case}
Here, we will suggest a concept of supersymmetric P-B oscillator.
For this aim, we should first take account of a set of algebraic
generators $\left\{ N, N^{^{\prime }}, Q, Q^{\dagger}, \sigma
_{3}\right\}$, which possesses a supersymmetric Lie algebraic
properties, {\it i.e.},

\begin{eqnarray}
&& Q^{2}=(Q^{\dagger })^{2}=0,   \quad  \left[
Q,Q^{\dagger}\right] =N^{^{\prime }}\sigma _{3},  \quad  \left[
N,N^{^{\prime }}\right] =0,   \nonumber \\
&&
\left[ N,Q\right] =-Q, \quad    \left[ N,Q^{\dagger }\right] =Q^{\dagger },    \nonumber \\
&&     \left\{ Q,Q^{\dagger }\right\} =N^{^{\prime }}, \quad
\left\{ Q,\sigma _{3}\right\} =\left\{ Q^{\dagger },\sigma
_{3}\right\} =0,    \nonumber \\
&&
\left[ N^{^{\prime }},Q\right] =0, \quad    \left[ N^{^{\prime }}, Q^{\dagger }\right] =0,    \nonumber \\
&& \left[ Q,\sigma _{3}\right]=-2Q,      \quad
 \left[ Q^{\dagger },\sigma _{3}
\right] =2Q^{\dagger },              \nonumber \\
&&               \left( Q^{\dagger }-Q\right) ^{2}=-N^{^{\prime
}},     \quad       \left[ N^{^{\prime }}, \sigma _{3} \right] =0,
\label{eq33}
\end{eqnarray}
where $\left\{\ \right\} $ denotes the anticommuting bracket. One
of the physical realization of such a Lie algebra is the
multiphoton Jaynes-Cummings model\cite{Sukumar,Kien}, the
generators of the interaction Hamiltonian of which is
$Q^{\dagger}=\frac{1}{\sqrt{k!}}{a^{\dagger}}^{k}\sigma_{-}$ and
$Q=\frac{1}{\sqrt{k!}}{a}^{k}\sigma_{+}$, where
$\sigma_{\pm}=\left(\sigma_{1}\pm i\sigma_{2}\right)/2$, and $k$
denotes the photon number in each atomic transition
process\cite{Sukumar,Kien,Shen2}. It can be readily verified that
the eigenvalue equation of the invariant operator $N^{^{\prime }}$
(which commutes with all the operators $N, Q^{\dagger}, Q, \sigma
_{3}$) is of the form
\begin{equation}
N^{^{\prime }} \left( {\begin{array}{*{20}c}
   {\left| m\right\rangle}  \\
   {\left| m+k\right\rangle}  \\
\end{array}} \right) =C_{m+k}^{m}\left( {\begin{array}{*{20}c}
   {\left| m\right\rangle}  \\
   {\left| m+k\right\rangle}  \\
\end{array}} \right)  \label{eq35}
\end{equation}
with the eigenvalue $C_{m+k}^{m}=\frac{(m+k)!}{m!k!}$. Further
calculation shows that
\begin{equation}
\left\{
\begin{array}{ll}
&         Q^{\dagger }\left( {\begin{array}{*{20}c}
   {\left| m\right\rangle}  \\
   {0}  \\
\end{array}} \right)=\sqrt{\frac{(m+k)!}{m!k!}}\left( {\begin{array}{*{20}c}
   {0}  \\
   {\left| m+k\right\rangle}  \\
\end{array}} \right),      \\
&         Q\left( {\begin{array}{*{20}c}
   {0}  \\
   {\left| m+k\right\rangle}  \\
\end{array}} \right)=\sqrt{\frac{(m+k)!}{m!k!}}\left( {\begin{array}{*{20}c}
   {\left| m\right\rangle}  \\
   {0}  \\
\end{array}} \right).
\end{array}
\right.
\end{equation}
Thus we obtain the following supersymmetric quasialgebra $\left\{
N, Q^{\dagger }, Q, \sigma _{3} \right\}$ in a sub-Hilbert-space
corresponding to the particular eigenvalue $C_{m+k}^{m}$ of the
invariant operator $N^{^{\prime }}$ by replacing the generator
$N^{^{\prime }}$ with $C_{m+k}^{m}$ in the commutation relations
in (\ref{eq33}), {\it i.e.},
\begin{eqnarray}
& &      \left[ Q,Q^{\dagger }\right] =C_{m+k}^{m}\sigma _{3},
\quad    \left\{ Q,Q^{\dagger }\right\} =C_{m+k}^{m},        \nonumber \\
& &      \left( Q^{\dagger }-Q\right) ^{2}=-C_{m+k}^{m}.
\label{eq36}
\end{eqnarray}

Here we assume that the supersymmetric P-B oscillator can be
characterized by the above supersymmetric quasialgebra
(\ref{eq36}). By analogy with the Hamiltonian
($H=\frac{1}{2}\left\{a,a^{\dagger }\right\}\omega$) of the
Bosonic oscillator, the Hamiltonian of such a supersymmetric P-B
oscillator may be written in the form $H=\frac{1}{2}\left\{
Q,Q^{\dagger }\right\}\Omega$, where $Q^{\dagger}$ and $Q$ can be
regarded as the creation and annihilation operators and the
eigenvalue $C_{m+k}^{m}$ of $N^{^{\prime }}$ may be considered the
particle occupation number of the supersymmetric P-B oscillator in
a certain number state ({\it e.g.}, the eigenstate of $N^{^{\prime
}}$).

It should be noted that the multiphoton Jaynes-Cummings model, the
Hamiltonian of which possesses a supersymmetric Lie algebraic
structure, can model the behavior of the supersymmetric P-B
oscillator: specifically, by choosing the appropriate parameters
for the multiphoton Jaynes-Cummings model, the Hamiltonian of
which can take the form of the linear combination of $Q$ and
$Q^{\dagger}$, namely, $H=gQ+g^{\ast}Q^{\dagger}$, where $g$ and
$g^{\ast}$ denote the coupling coefficient of this Jaynes-Cummings
model. If the stationary Schr\"{o}dinger equation of the
multiphoton Jaynes-Cummings model is of the form
$H|\psi\rangle=\epsilon |\psi\rangle$, then one can easily obtain
a new eigenvalue equation $\left\{ Q,Q^{\dagger
}\right\}|\psi\rangle=\epsilon^{2}/(g^{\ast}g)|\psi\rangle$.

\section{A potential application}
In the previous sections, we discussed the Lie algebraic structure
of P-B oscillator and then proposed a supersymmetric
generalization. Here, we will consider a potential application of
the concept of the supersymmetric P-B oscillator to the mass
spectrum of charged leptons. First, we suggest a new mass formula
for the charged leptons that agrees with experimental values to a
high degree of accuracy. This leptonic mass spectrum is
constructed based on the following three clues: (i) the
supersymmetric P-B oscillator; (ii) Barut's viewpoint of magnetic
self-interaction of charged leptons\cite{Barut1,Barut2}; (iii) the
experimental values of charged leptons.

According to the previous section, the energy (or frequency) of
the supersymmetric P-B oscillator is proportional to the
combination coefficient $C^{l}_{m}$ ({\it i.e.},
$\frac{m!}{l!(m-l)!}$). Particularly, if the two integers $m$ and
$l$ satisfy the relation $m=l$, then this supersymmetric
oscillator will be reduced to a regular two-dimensional
Pegg-Barnett oscillator. We assume that there is a deep connection
between the supersymmetric Pegg-Barnett oscillator and the
generation replication of charged leptons, namely, the
supersymmetric Pegg-Barnett oscillator can model some behaviors
(at least the mass spectrum) and the internal structures (should
such exist) of some certain elementary particles. Thus, the above
combination coefficient should be introduced into the leptonic
mass spectrum under consideration. In addition, Barut showed that
the mass difference between muon and electron may result from the
magnetic self-interaction energy of the
electron\cite{Barut1,Barut2}. He believed that the radiative
effects give an anomalous magnetic moment to the electron, which
implies an extra magnetic energy\cite{Barut1,Barut2}. By using the
quantization formulation according to the Bohr-Sommerfeld
procedure, one can obtain the magnetic energy of a system
consisting of both a charge and a magnetic moment as
$E_{l}=\lambda l^{4}$ with $l$ and $\lambda$ being the angular
quantum number of the system and a certain constant
coefficient\cite{Barut1,Barut2}. Based on the above enlightening
clues, a new mass formula for the charged leptons can be
constructed as follows
\begin{equation}
m_{n}=\left(1+\frac{1}{2\alpha}\sum^{n}_{l=0}C^{l}_{3}l^{4}\right)m_{\rm
e},    \label{formula}
\end{equation}
where $m_{\rm e}$, $\alpha$ and $n$ denote the electron mass, the
electromagnetic fine structure constant and the generation quantum
number (generation index of leptons), respectively. Here, the
electron (e), muon ($\mu$) and tau ($\tau$) particle correspond to
$n=0, 1, 2$, respectively. By using $\alpha^{-1}=137.036$ and
$m_{\rm e}=0.51100$ MeV, it follows from the formula
(\ref{formula}) that the masses of charged leptons of various
generations are $m_{\mu}=105.55$ MeV, $m_{\tau}=1786.2$ MeV and
$m_{3}=4622.2$ MeV. The experimental values for muon and tau
masses are $m_{\mu}^{\rm exp}=105.66$ MeV, $m_{\tau}^{\rm
exp}=1784.2$ MeV\cite{Stone}. Thus the relative precisions of the
formula (\ref{formula}) are only of $-1.04\times 10^{-3}$ (for
muon) and $+1.12\times 10^{-3}$ (for tau), respectively. It should
be noted that the currently accepted value for the tau lepton mass
that was measured in 1992 is $m_{\tau}^{\rm exp}=1776.9$
MeV\cite{Bai}. Someone may therefore argue that the formula
(\ref{formula}) will not agree very well with such an experimental
result. But we will point out that this is not the true case.
Since for the case of tau, the interaction energy scale is of GeV,
one should consider the running coupling ``constant" of
electromagnetic interaction ( {\it i.e.}, the variation of the
fine structure constant at different energy scale). By taking
account of this factor, the inverse of the fine structure
constant, $\alpha^{-1}$, will decrease. As a result, the mass of
tau particle obtained by using (\ref{formula}) will therefore
become less and is still in good agreement with the experimental
value obtained in 1992\cite{Bai}. In a word, the mass spectrum of
charged leptons presented in this paper can agree with the
experiments to about one part in $10^{3}$. It is shown from the
mass spectrum (\ref{formula}) that there may exist a fourth (and
even final) charged lepton, the mass ($m_{3}$) of which is more
than 9000 times that of electron. Even though such a heavy
``electron" has so far never been observed experimentally, such a
mass spectrum may still be of interest, since the most remarkable
feature of (\ref{formula}) is that the total generation number of
leptons is {\it finite} (rather than infinite), which results from
the fact that the integer $l$ in the combination number
$C^{l}_{3}$ can be taken to be $l=0, 1, 2, 3$ only. So, it might
be able to interpret one of the most fundamental problems in
particle physics and quantum field theory, {\it i.e.}, the
finite-generation-number phenomenon of fermions. This advantage
has never arisen in the previous mass spectra of leptons proposed
by other investigators, where the mass formulae could not suppress
the generation index and the total number of the allowed
generations of the fermion chain is
infinite\cite{Barut1,Barut2,Tennakone}.

It should be noted that although the present experimental
evidences show that the total generation number of the fermion
chain is three\cite{Dolgov}, a probe into the potential existence
of extra generations of fermions has still attracted attention of
many investigators by
now\cite{Arik1,Arik2,Lemke,Iwao,Mohapatra,Kniehl,Huang,Novikov,Novikov2,Belotsky,Alan}.
In experiments, the latest electroweak precision data allows the
existence of additional chiral generations in the standard
model\cite{Arik2}. Arik {\it et al.} studied the influence of
extra generations on the production of the standard model Higgs
boson at hadron colliders\cite{Arik1,Arik2}. In theoretical
investigations, some authors extended the standard electroweak
gauge model to include a fourth generation of fermions, and
considered the exotic interactions involving fourth-generation
quarks and leptons which cannot be confused experimentally with
those of the standard model, or suggested a completely different
interaction model for the extra-generation
fermions\cite{Huang,Novikov,Novikov2,Belotsky,Alan,Cheng,Ma,Datta,Huo,Solmaz}.
These studies may provide a possible test of the fourth generation
and would give a signal of new physics.

\section{Conclusion}
We discuss the su$(n)$ Lie algebraic structure of the P-B
oscillator and generalize it to a supersymmetric case. We think
that in some sense the multiphoton Jaynes-Cummings model can
describe the behavior of the supersymmetric P-B oscillator. Based
on the concept of supersymmetric P-B oscillator and Barut's
viewpoint of magnetic self-interaction of charged leptons, we
construct a new mass formula for charged leptons, the most
remarkable feature of which is such that the total number of the
generations of the charged leptons is finite. We think that since
it can present a possible explanation for the finite-generation
phenomenon of charged leptons, our tentative analysis (the
application of supersymmetric P-B oscillator to leptonic mass
spectrum) in the present paper may still deserve further
consideration. We hope this consideration might provide us with an
insight into the problems such as the physical origin of the
generation replication of the fermion chain.

\textbf{Acknowledgements}  This work was supported partially by
the National Natural Science Foundation of China under Project No.
$90101024$.

\end{document}